\documentstyle[epsf,psfig,rotate]{l-aa}

\def\htre{~$h^{-3}$ Mpc$^3$~}

\def\kms{km s$^{-1}$\ }
\def\drho{$\delta\rho/\rho$\ }
\def\mpc{$h^{-1}$ Mpc}
\def\v0{V$_0$\ }
\def\d0{D$_0$\ }
\def\vl{V$_L$\ }
\def\dl{D$_L$\ }
\def\cz{$cz$\ }
\def\sv{$\sigma_{cz}$\ }
\def\nmem{$N_{mem}$\ }

\def\egals{$e.l.$-galaxies \ }

\def\agals{$a.l.$-galaxies \ }
\def\htre{~$h^{-3}$ Mpc$^3$~}
\def\etal{{et al.}\ } 
\newbox\grsign \setbox\grsign=\hbox{$>$} \newdimen\grdimen \grdimen=\ht\grsign
\newbox\simlessbox \newbox\simgreatbox
\setbox\simgreatbox=\hbox{\raise.5ex\hbox{$>$}\llap
     {\lower.5ex\hbox{$\sim$}}}\ht1=\grdimen\dp1=0pt
\setbox\simlessbox=\hbox{\raise.5ex\hbox{$<$}\llap
     {\lower.5ex\hbox{$\sim$}}}\ht2=\grdimen\dp2=0pt

\def\simless{\mathrel{\copy\simlessbox}}
\begin{document}
\thesaurus{ 11(11.03.1; 11.04.01; 11.12.2; 12.03.3; 12.12.1) }
\title{ The ESO Slice Project (ESP) galaxy redshift survey:
\thanks{based on observations collected at the European Southern
Observatory, La Silla, Chile.} }
\subtitle{VI. Groups  of Galaxies}
%
\author{
M.Ramella \inst{1}
\and
G.Zamorani \inst{2,3}
\and
E.Zucca \inst{2,3}
\and
G.M.Stirpe \inst{2}
\and
G.Vettolani \inst{3}
\and
C.Balkowski \inst{4}
\and
A.Blanchard \inst{5}
\and
A.Cappi \inst{2}
\and
V.Cayatte \inst{4}
\and
G.Chincarini \inst{6,7}
\and
C.Collins \inst{8}
\and
L.Guzzo \inst{6}
\and
H.MacGillivray \inst{9}
\and
D.Maccagni \inst{10}
\and
S.Maurogordato \inst{11,4}
\and
R.Merighi \inst{2}
\and
M.Mignoli \inst{2}
\and
A.Pisani \inst{1}
\and
D.Proust \inst{4}
\and
R.Scaramella \inst{12}
}
%
\institute{ 
Osservatorio Astronomico di Trieste, 
via Tiepolo 11, 34131 Trieste, Italy
\and
Osservatorio Astronomico di Bologna, 
via Zamboni 33, 40126 Bologna, Italy
\and
Istituto di Radioastronomia del CNR, 
via Gobetti 101, 40129 Bologna, Italy
\and
Observatoire de Paris, DAEC, Unit\'e associ\'ee au CNRS, D0173 et \`a
l'Universit\'e Paris 7, 5 Place J.Janssen, 92195 Meudon, France
\and
Universit\'e L. Pasteur, Observatoire Astronomique, 
11 rue de l'Universit\'e, 67000 Strasbourg, France
\and
Osservatorio Astronomico di Brera, 
via Bianchi 46, 22055 Merate (LC), Italy
\and
Universit\`a degli Studi di Milano, 
via Celoria 16, 20133 Milano, Italy
\and
Astrophysics Research Institute, Liverpool John--Moores University, 
Byrom Street, Liverpool L3 3AF, United Kingdom
\and
Royal Observatory Edinburgh, 
Blackford Hill, Edinburgh EH9 3HJ, United Kingdom
\and
Istituto di Fisica Cosmica e Tecnologie Relative, 
via Bassini 15, 20133 Milano, Italy
\and
CERGA, Observatoire de la C\^ote d'Azur, 
B.P. 229, 06304 Nice Cedex 4, France
\and
Osservatorio Astronomico di Roma, 
via Osservatorio 2, 00040 Monteporzio Catone (RM), Italy
}
%
%
\offprints{Massimo Ramella (ramella@oat.ts.astro.it)}
\date{Received 00 - 00 - 0000; accepted 00 - 00 - 0000}
\maketitle
\markboth {M.Ramella et al.: 
The ESP galaxy redshift survey: VI. Groups of Galaxies}{}
\begin{abstract}
In this paper we identify objectively and analyze groups of 
galaxies in the recently completed ESP survey 
($23^{h} 23^m \le \alpha_{1950} \le 01^{h} 20^m $ and 
$22^{h} 30^m \le \alpha_{1950} \le 
22^{h} 52^m $; $ -40^o 45' \le \delta_{1950} \le -39^o 45'$).  
We find 231 groups above the number overdensity threshold \drho=80
in the redshift range 5000 \kms $\le cz \le $ 60000 \kms.
These groups contain 1250 members, 40.5 \% of the 3085 ESP
galaxies within the same \cz range. 

The median
velocity dispersion (corrected for measurement errors and computed
at the redshift of the group) is $\sigma_{ESP,median}$ =
194 \kms. We show that our result is reliable in spite of the
particular geometry of the ESP survey (two rows of tangent circular
fields of radius $\theta = 15$ arcmin), which causes most systems to
be only partially surveyed. In general, we find that
the properties of ESP groups are consistent with those of groups
in shallower (and wider) catalogs  (e.g. CfA2N and SSRS2).
As in shallower catalogs, ESP groups trace very well 
the geometry of the large scale structure.
Our results are of particular interest 
because the depth of the ESP survey allows us
to sample group properties over a large number of structures.

We also compare luminosity function and spectral properties of
galaxies that are members of groups with those of isolated galaxies.
We find that galaxies in groups have a brighter $M^*$ with respect to
non--member galaxies; the slope $\alpha$ is the same, within the errors,
in the two cases.  We find that 34\% (467/1360) of ESP galaxies with
detectable emission lines are members of groups. The fraction of
galaxies without detectable emission lines in groups is significantly
higher: 45\% (783/1725).  More generally, we find  a  gradual decrease of the
fraction of emission line galaxies among members of systems of
increasing richness. This result confirms that the morphology-density
relation found for clusters also extends toward systems of lower
density.  
\footnote{Table 1 is available only in electronic form via anonymous 
ftp to cdsarc.u-strasbg.fr (130.79.128.5)
or via http://cdsweb.u-strasbg.fr/Abstract.html}

\keywords{Galaxies: clusters: general - distances and redshifts - luminosity
function, mass function;
          Cosmology: observations - large--scale structure of the Universe }
\end{abstract}

\section{Introduction}

The study of groups of galaxies as dynamical systems is interesting not
only {\it per se}, but  also because groups can be used to set constraints
on  cosmological models (e.g. Frenk \etal 1990; Weinberg \& Cole 1992,
Zabludoff \etal 1993;
Zabludoff \& Geller 1994, Nolthenius \etal 1994, 1997) and on models of
galaxy formation (Frenk \etal 1996; Kaufmann \etal 1997).  Groups are
also interesting sites where to look for interactions of galaxies with
their environment, in order to obtain information on galaxy evolution
processes (Postman \& Geller 1984, Allington-Smith \etal 1993).

Group catalogs identified in redshift space are increasing both in
number and size (CfA2N, RPG; SSRS2, Ramella \etal 1998;
Perseus-Pisces, Trasarti Battistoni 1998; LCRS, Tucker \etal 1997). At the same time, cosmological n-body simulations are
reaching the resolution required to allow replication of the
observational techniques for the identification of groups.  In
particular, Frederic (1995a,b) uses n-body simulations to  evaluate and
compare the performances of commonly used group finding algorithms.

Among the main properties of groups, the velocity dispersion
is of particular interest. It is easy to measure and it is
well suited for comparison with the predictions  of cosmological n-body
models (Frenk \etal  1990; Moore \etal 1993; Zabludoff \etal 1993). 
Distributions of velocity dispersions of nearby groups are now well 
determined with rather small statistical uncertainties given the
large size of the samples. Ramella \etal 1995 and Ramella \etal 1996 
survey the redshifts of candidate faint members of a 
selection of nearby groups and 
find that the velocity dispersion 
of groups is stable against  inclusion of 
fainter members. In other words, the velocity dispersion estimated 
on the basis of the fewer original bright members is a good indicator of the 
velocity dispersion obtained from a better sampling of the same group.

In this paper we identify and analyze groups of galaxies in
the recently completed ESP survey (Vettolani \etal 1997). 
The ESP group catalog is interesting because of 
its depth ($b_J \le 19.4$) 
and because it samples a new independent region of the universe.
ESP is a nearly bi-dimensional survey (the declination range is
much smaller than the right ascension range), five times deeper than either
CfA2 (Geller \& Huchra 1989) or SSRS2 (da Costa \etal 1994). 
The volume of the survey is $\sim 5 \times 10^4$ \htre at 
the sensitivity peak of the survey,
$z \sim 0.1$, and $\sim 1.9\times 10^5$ 
\htre at the effective depth of the sample, $z \sim 0.16$.
Even if the volume of ESP is of the 
same order of magnitude of the
volumes explored with the individual CfA2, SSRS2, and Perseus-Pisces samples, 
it intercepts a larger number of structures. In fact, the strip geometry
is very efficient for the detection of large scale structures 
within redshift surveys (de Lapparent \etal 1988). 

In particular we determine the distribution of the velocity dispersions
of groups and show that our result is reliable in spite of the
particular geometry of the ESP survey (two rows of adjacent circular
fields of radius $\theta = 16$ arcmin, see Figure 1 of  
Vettolani \etal 1997).

An important aspect that distinguishes the ESP group catalog from  the
other shallower catalogs is that we have the spectra of all galaxies with
measured redshift.
It is already well known that emission line galaxies
are rarer in rich clusters than in the field (Biviano \etal 1997).  The
relation between the fraction of emission line galaxies and the local
density is a manifestation of the morphology--density relationship
observed for clusters (Dressler 1980), a useful tool in the study of galaxy evolution.
With the ESP catalog we explore the extent of the morphology density
relationship in the intermediate range of densities that are typical of
groups at a larger depth than in previous studies.

We note that  preliminary
results of a search for groups in the Las Campanas Redshift Survey 
(Shectman \etal 1996) have
been presented by Tucker \etal (1997).  The properties of these groups,
as distant as ours, are difficult to compare with those of our ESP
groups and with those of shallower surveys because LCRS  a) is a red
band survey (ESP and shallower surveys are selected
in the blue band), b) it is
not simply magnitude limited, and c) it does not uniformly sample
structures in redshift space. In particular, the different 
selection criteria could have a strong impact
on the results concerning the morphology-density relation,
the luminosity segregation, and the possible differences between the 
luminosity functions of member and non-member galaxies. 

In section 2) we briefly describe the data; in section 3) we analyze 
the effect of the ESP geometry on the estimate of the velocity dispersion
of groups; in section 4) 
we summarize the group identification procedure; in section 5) we present
the ESP group catalog; in section 6) we analyze properties of groups
that are relevant to a characterization of the Large Scale Structure (LSS); 
in section 7) we analyze the properties of
galaxies in groups and compare them to the properties of ``field''
galaxies ({\it i.e. } galaxies that have not been assigned to groups) and ``cluster'' galaxies; in section 8) we identify ESP counterparts
of ACO and/or EDCC clusters (Lumsden \etal 1992). 
Finally, we summarize our results in section 9).
\section{The Data}

The ESO Slice Project (ESP) galaxy redshift survey is described 
in Vettolani \etal (1997). The data of the full sample together
with a detailed description of the instrumental set-up and of the
data reduction can be found in Vettolani \etal (1998). Here
we only briefly describe the survey.
 
The ESP survey extends over a strip of $\alpha \times \delta = 22^o \times 
1^o$ (strip A), plus a nearby area of $5^o \times 1^o$ (strip B), five degrees
West of the main strip, in the South Galactic Pole region (
$23^{h} 23^m \le \alpha_{1950} \le 01^{h} 20^m $ and 
$22^{h} 30^m \le \alpha_{1950} \le 
22^{h} 52^m $ respectively; $ -40^o 45' \le \delta_{1950} \le -39^o 45'$). 
Each of the two strips is covered with two rows
of slightly overlapping circular fields of angular radius $\theta =
16$ arcmin, the separation between the centers of neighboring
circles being 30 arcmin. 
Each field corresponds to the field of view of the multifiber
spectrograph OPTOPUS at the 3.6m ESO telescope that was used
to obtain almost all of the spectra (the MEFOS spectrograph
was used in the last ESP run).
Throughout this paper we will assume that the circular fields are
tangent, with an angular radius of 15 arcmin: 
this simplification has no consequences on the galaxy sample. 
The total solid angle of the spectroscopic survey is 23.2 square degrees.

The galaxy catalog consists of all (candidate) galaxies 
brighter than the limiting magnitude $b_{J,lim} = 19.4$ listed 
in the Edinburgh--Durham Southern Galaxy Catalogue (Heydon--Dumbleton 
et al. 1988, 1989).
 
The spectra cover the wavelength range 3730\AA~to 6050\AA, with 
an average pixel size of 4.5\AA. 
Galaxy redshifts are measured by cross--correlating 
sky-subtracted spectra with a set of 8 
template stars observed with 
the same instrumental set-up used to obtain the galaxy spectra. 
In this paper we use emission line redshifts only for
galaxies with no reliable absorption line redshift.
The median internal velocity error is of the order of $\sim 60$ \kms.
From a comparison of our 8 templates with three SAO radial velocity
standard stars we estimate that the zero--point error should be smaller
than $\sim 10$ \kms. 

The total number of confirmed galaxies with reliable redshift measurement
is 3342. The completeness of strip A and strip B are estimated to be
91\% and 67\% respectively. 

\section{ESP Geometry and the Measure of Velocity Dispersions}

To all practical purposes, the projection of the ESP survey on the sky
consists of two rows of adjacent
circular OPTOPUS fields of radius 15 arcmin and a separation
of 30 arcmin between adjacent centers.  The
angular extent of groups and clusters at the typical depth of the
survey ($z \simeq 0.1$ ) are comparable, or even larger, than the size
of the OPTOPUS fields. Therefore, most systems falling into the
survey's volume are only partially surveyed.

The main effect of the "mask" of OPTOPUS fields is to hide a fraction
of group members that lie within or close to the strip containing the
mask (the OPTOPUS fields cover 78\% of the area of the "un-masked"
strip).  Because of the hidden members, several poor groups may not
appear at all in our catalog.  On the contrary, our catalog might
include parts of groups that are centered outside the ESP strip.  These
problems notwithstanding, we expect to derive useful information on the
most important physical parameter of groups, the velocity dispersion,
\sv.

Our estimate of 
the parent velocity dispersion, $\sigma_p$, is based upon
the sample standard deviation $\sigma_{cz}(N_{mem})$. The sample standard
deviation defined as $ \sigma_{cz}(N_{mem}) = \sqrt{\Sigma _i (cz_i - <cz>)^2/
(N_{mem}-3/2)}$ is a nearly unbiased estimate of the velocity
dispersion (Ledermann, 1984), independent of the size $N_{mem}$ of
the sample. We make the standard assumptions that
a) barycentric velocities of members are not correlated with
their real 3D positions within groups, and that b) 
in each group the distribution of barycentric 
velocities is approximately
gaussian. Because the position on the sky of the OPTOPUS mask is not
related to the positions of groups, its only effect is to
reduce at random $N_{mem}$. Therefore, using an unbiased estimate
of the velocity dispersion, the mask
has no effect on our determination of the average velocity dispersions of
groups.

The effect of the mask is to broaden the distribution of the sample
standard deviations. The variance of the distribution of sample
standard deviations varies with $N_{mem}$ approximately as
$\sigma_{cz}^2/2N_{mem}$ (Ledermann, 1984).  
This distribution, proportional to the
$\chi^2$ distribution, is skewed: even if the mean of the distribution
is unbiased, \sv is more frequently underestimated than overestimated.

While it is easy to predict the effect of the mask on the determination
of the velocity dispersion of a single group, it is rather difficult to
predict the effect of the mask on the observed distribution of velocity
dispersions of a sample of groups with different ``true'' velocity
dispersions and different number of members.  In order to estimate
qualitatively the effect of the mask on the shape of the distribution
of velocity dispersions, we perform a simple Monte Carlo simulation.

We simulate a group by placing uniformly at random $N_{mem}$ points
within a circle of angular radius $\theta_{gr}$ corresponding, at the
redshift of the group, to the linear projected radius $R_{gr}$ = 0.5
\mpc. This radius is the typical size of groups observed in shallow
surveys (e.g. RPG).  We select the redshift of the
group, $z_{gr}$, by random sampling the observed distribution of ESP
galaxy redshifts.

In order to start from reasonably realistic distributions, 
we set $N_{mem}$ and the velocity dispersion, \sv, by random sampling the
relative histograms obtained from our ESP catalog. We limit the range
of \nmem to 3 $\le$ \nmem  $\le $ 18 and the range of \sv to 0 $\le $
\sv $\le$ 1000 \kms.

We lay down at random the center of the simulated group within the region
of the sky defined by extending 15 arcmin northward and southward the
"un-masked" limits of the ESP survey.  We then assign to each of the
\nmem points a barycentric velocity randomly sampled from a gaussian
with dispersion \sv centered on $z_{gr}$.  We compute the velocity
dispersion, $\sigma_{no-mask}$, of the \nmem velocities.  Finally, we
apply the mask and discard the points that fall outside the mask.  
We discard the whole group if there
are fewer than 3 points left within the mask ($N_{mem,mask} < 3$). If
the group ``survives'' the mask, we compute the dispersion
$\sigma_{mask}$ of the $N_{mem,mask}$ members. On average, 78\% of the
groups survive the mask (this fraction corresponds to the ratio between
the area covered by the mask and the area of the ``un-masked'' strip). 
The percentage of surviving groups depends on
the exact limits of the region where we lay down at random groups and
on the projected distribution of members within $R_{gr}$. For the
purpose of the simulation, the fraction of surviving groups is not
critical.

We repeat the procedure 100 times for $n_{gr}$ = 231 simulated groups
(the number of groups identified within ESP).  At each run we compute
the histograms N($N_{mem,mask}$) and N($\sigma_{mask}$).

In Figure 1 we plot the input distribution N($N_{mem}$) --thin line--
together with the average output distribution, 
$<$N($N_{mem,mask}$)$> n_{gr}/n_{mask}$
-- thick line--.  Errorbars
represent $\pm$ one standard deviation derived in each bin from the
distribution of the 100 histograms N($N_{mem,mask}$); for clarity we omit 
the similar errorbars of N($N_{mem}$.  
The factor $n_{gr}/n_{mask}$ normalizes the output distribution
to the number of input groups. The two
histograms in Figure 1 are clearly very similar since N($N_{mem}$) is
within one sigma from $<$N($N_{mem,mask}$)$> n_{gr}/n_{mask}$ for 
all values of \nmem. We point out here that the similarity between the input and
output histograms does not mean that the surviving groups have not
changed. In fact only about 63\% of the triplets survive the mask
while, for example,  88\% of the groups with 5 members and 98\%  of
those with 10 members ``survive''  the mask.

\begin{figure}
\epsfysize=9cm
\epsfbox{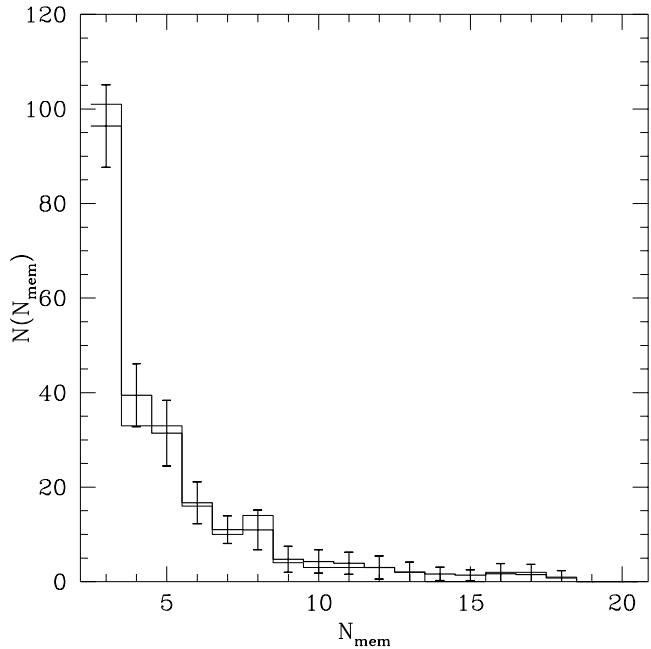}
\caption[]{Effect of the ``OPTOPUS mask'' on N($N_{mem}$): 
the thin histogram is the ``true'' distribution,
the thick histogram is the average distribution ``observed'' through the
``OPTOPUS mask'', normalized to the number of input groups.  
Errorbars represent $\pm$ one standard deviation.
}
\end{figure}

Figure 2 shows the results of our simple simulation for the velocity
dispersion. The thin histogram is the input ``true'' distribution
N($\sigma_{gr}$). The dotted histogram is
the average ``observed'' distribution obtained without 
dropping galaxies that lie outside the OPTOPUS mask, N($\sigma_{no-mask}$).
This is the distribution we would observe if the
geometry of the survey would be a simple strip.
The third histogram (thick line) is the average output
distribution in presence of the OPTOPUS mask, 
$<$N($\sigma_{mask}$)$> n_{gr}/n_{mask}$  
(errorbars are $\pm$  one-sigma).   The input
distribution N($\sigma_{gr}$), the distribution N($\sigma_{no-mask}$),
and the distribution $<$N($\sigma_{mask}$)$> n_{gr}/n_{mask}$
are all within one-sigma from each other.
In particular the two distributions we observe with and without
OPTOPUS mask are undistinguishable (at the 99.9\% confidence
level, according to the KS test). The low velocity dispersion
bins are slightly more populated in the ``observed'' histograms 
because the estimate of the ``true'' \sv is based on small \nmem.
Note that in the case of real observations, some 
groups in the lowest \sv bin will be shifted again to the next higher bin
because of  measurement errors.

Our results do not change if we take into account the slight dependence
of \sv from \nmem observed within our ESP catalog:  also in this case
the effect of the mask is negligible.

In conclusion, the simulation confirms our expectation that the OPTOPUS
mask has no significant effect on the shape of the distribution of
velocity dispersions.
\begin{figure}
\epsfysize=9cm
\epsfbox{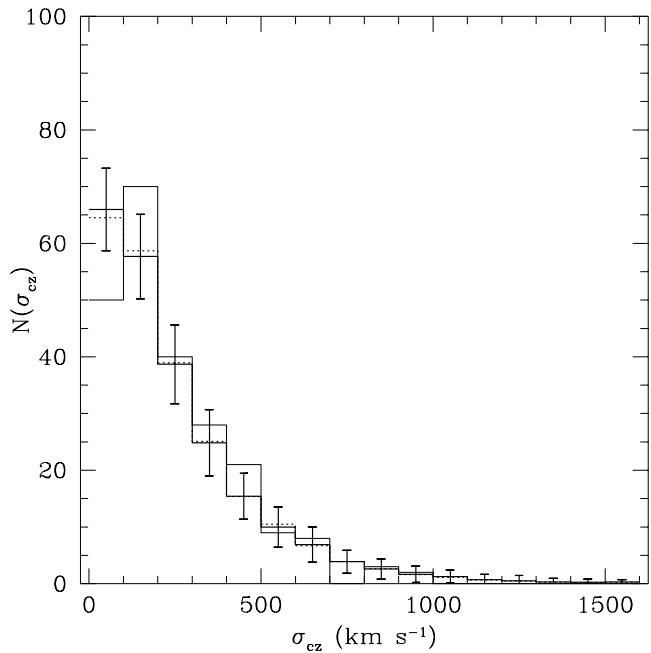}
\caption[]{Effect of the ``OPTOPUS mask'' on N($\sigma_{gr}$). 
The thin histogram is the ``real'' distribution, the dotted 
histogram shows the effect of
sampling on the input distribution, and
the thick histogram is the average  distribution  
``observed'' through the ``OPTOPUS mask'', 
normalized to the number of input groups. 
Errorbars represent $\pm$ one standard deviation.
}
\end{figure}

\medskip \medskip

\section{Group Identification}

We identify groups with the so-called friend-of-friend algorithm (FOFA;
Huchra \& Geller, 1982) as described in RPG. 
We implement here the cosmological corrections required
by the depth of the sample ($z \le 0.2$). Throughout this paper we
use H$_o$ = 100 \kms Mpc$^{-1}$ and $q_0 = 0.5$.

For each galaxy in the magnitude limited ESP catalog, 
the FOFA identifies all other galaxies
with a projected comoving separation $$D_{12} \le  D_L(V_1,V_2) \eqno (1)$$ 
and a line-of-sight velocity difference

$$V_{12} \le V_L(V_1,V_2). \eqno (2)$$

Here $V_1 = cz_1$ and $V_2 = cz_2$ are the velocities of the 
two galaxies in the pair. 
All pairs linked by a common galaxy form a ``group''.

The two parameters \dl, \vl are scaled with distance in order to take
into account the decrease of the magnitude range of the luminosity
function sampled at increasing distance. The scaling is
 
$$D_L=D_o R \eqno (3)$$
 
and
 
$$V_L = V_o R, \eqno (4)$$
 
where

$$R=\left[\int_{-\infty}^{M_{lim}} \Phi(M)
{dM}/\int_{-\infty}^{M_{12}} \Phi(M) {dM}\right]^{1/3}, \eqno (5)$$

$$M_{12}= b_{J,lim}-25-5 \log(d_{L}(\bar{z}_{12})) - <K(\bar{z}_{12}>, \eqno (6) $$

and $M_{lim}$ is the absolute magnitude corresponding to $b_{J,lim}$
at a fiducial velocity $V_f$. We
compute $d_L(\bar{z}_{12})$ with the Mattig (1958) expression, where  
$\bar{z}_{12} = .5(z_1 + z_2)$. Finally, $<K(\bar{z}_{12})>$ is the 
$K-$correction ``weighted'' with the expected morphological mix at
each redshift as in Zucca \etal (1997).

The scaling is the same for both \dl and \vl and it is normalized 
at the fiducial velocity  $V_f$ = 5000 \kms, 
where \d0 = $D_L(V_f$) and \v0 = $V_L(V_f$). In particular,
a given value of \d0 corresponds to a  minimum number overdensity
threshold for groups, \drho.  The luminosity function we use is the
Schechter parametrization with $M^* = -19.61$, 
$\alpha = -1.22$, and $\phi^* = 0.020$ Mpc$^{-3}$
computed for ESP galaxies by Zucca \etal (1997). 

We do not consider galaxies with velocities
$cz \le V_f$ because the linear extension of the survey in the
direction of the declination is smaller than the typical size
of a group for $cz \le $5000 \kms. 
We also limit the maximum depth of our group catalog to $cz \le 60000$
\kms.  Beyond this limit the accessible part of the luminosity function
becomes very small and the scaling of the FOFA linking parameters
excessively large.

The main characteristics of the distribution of galaxies within the
volume of the universe surveyed by ESP (Vettolani \etal 1997) are very
similar to those observed within shallower, wider angle magnitude
limited redshift surveys.  For this reason we expect that the
conclusions on the fine-tuning of FOFA reached by Ramella \etal 1989,
Frederic 1995a, and RPG will hold true also for ESP.
In particular, RPG show that within the CfAN2 redshift survey the
choice of the FOFA parameters is not critical in a wide region of the
parameter space around (\drho = 80, \v0 = 350 \kms).  With our
luminosity function and fiducial velocity, we obtain \drho = 80 for \d0
= 0.26 Mpc, a value comparable to the \d0 value used for CfAN2. It is
therefore reasonable to expect that the same results of the exploration
of the parameter space will hold also for the ESP survey. In order to
verify our expectation, we run FOFA with the following five pairs of
values of the linking parameters selected among those used by RPG:
(\drho = 80, \v0 = 250 \kms), (\drho = 80, \v0 = 350 \kms), (\drho =
80, \v0 = 600 \kms), (\drho = 60, \v0 = 350 \kms), (\drho = 100, \v0 =
350 \kms). Based on RPG, these pairs of values are sufficient to give
an indication of the stability of the group catalogs in the parameter
space (\dl,\vl).  The number of groups in the five cases is
N$_{groups}$ = 217, 231, 253, 239, and 217 respectively.

\begin{figure}
\epsfysize=9cm
\epsfbox{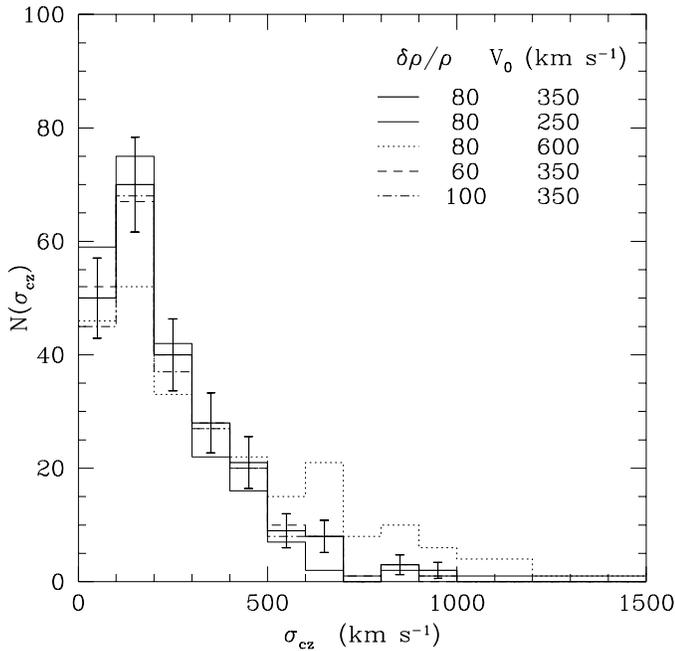}
\caption[]{The distribution of velocity dispersions, N($\sigma_{cz}$),
of different ESP group catalogs obtained for a grid of values of
the search parameters \drho and $V_0$.
Errorbars represent $\pm$ one standard deviation.
}
\end{figure}

We plot in Figure 3 the observed distributions of the velocity 
dispersions of the five group catalogs.
We compare the distribution obtained for (\drho = 80, \v0 = 350 \kms)
--thick histogram in Figure 3 -- with the other four 
distributions and find that the only significant
difference (99.9 \% level, according to the KS test) occurs with
the distribution obtained using the largest velocity link, \v0 = 600
\kms (dotted histogram).  This value of the 
velocity-link produces an excess of high
velocity dispersion systems.  Frederic 1995a, 
Ramella \etal 1989, and RPG argue that these high velocity 
dispersion systems are likely to
include a significant number of interlopers 
(galaxies with high barycentric velocity that are
not physically related to the group in real space).

On the basis of the results of our tests, we choose the catalog
obtained with  \drho = 80  (\d0 = 0.26 \mpc) and \v0 = 350 \kms as our
final ESP group catalog. This choice offers the advantage of a
straightforward comparison between the properties of ESP catalog and
those of the CfA2N (RPG), and SSRS2 (Ramella \etal 1998)
catalogs.
\section{The Group Catalog}

We identify 231 groups within the redshift limits $5000 \le cz \le
60000$ \kms. These groups contain 1250 members, 40.5 \% of the 3085 ESP
$b_J \leq 19.4$ galaxies within the same \cz range.

In Table 1 we present our group catalog. For each group we list the ID
number (column 1), the number of members (column 2), the coordinates
$\alpha_{1950}$ and $\delta_{1950}$ (columns 3 and 4 respectively), the
mean radial velocity \cz in \kms corrected for Virgo infall and
galactic rotation (column 5), and the velocity dispersion \sv (column 6).
We compute the velocity dispersion following the prescription of Ledermann
(1984) for an unbiased estimator of the dispersion (see previous section).
We also take into account the cosmological expansion of the universe
and the measurement errors according to the prescriptions of 
Danese \etal (1980).  
The errors we associate to the redshifts are those output by the RVSAO
cross-correlation procedure multiplied by a factor 1.6. This factor 
brings the cross-correlation error in rough agreement 
with the external error estimated from repeated observations (Vettolani
\etal 1998 -- here we do not distinguish between emission and
absorption line redshifts). 
Table 1 is available 
only in electronic form via anonymous ftp to cdsarc.u-strasbg.fr (130.79.128.5)
or via http://cdsweb.u-strasbg.fr/Abstract.html.

In the case of 24 groups, the correction of \sv for the measurement
errors leads to a negative value. In column 6 of Table 1 we give the error 
as an upper limit to \sv for these groups.

Not all galaxies in the region of the sky covered by the ESP survey
have a measured redshift. Of the original target list, 444 objects are
not observed, and 207 objects have a noisy spectrum, insufficient  for
a reliable determination of the redshift.  In Table 1 we give, for each
group, the ratio of these objects to the number of members (column 7).
In computing these rates, 
we assign to each group the objects without redshift whose ($\alpha$,
$\delta$) coordinates fall within the angular region defined by the
union of the \nmem circular regions obtained by projecting on the sky
circles of linear radius $D_L(cz_{group})$ centered on all group
members.  There are groups that are separated along the line-of-sight
but that overlap once projected on the sky.  If an object without
redshift lies within the overlap region, we assign the object to both
groups.

There are 67 groups that do not contain any object of the photometric
catalog without measured
redshift. On the other hand, in the case of 51 groups the number of objects without
redshift equals, or exceeds, the number of members.  These groups
are mostly triplets and quadruplets. Only 14 out of the 51 (possibly)
highly incomplete groups have \nmem $\geq$ 5. Most of these groups are located
in the relatively small region B of the redshift survey (Vettolani \etal 1998),
which is the least complete (completeness level = 71\%).

Finally, we estimate that only 8 out of 231 groups are entirely contained
within one OPTOPUS field. By "entirely contained" we mean that none of
the circles of projected linear radius  $D_L$ centered on the member
galaxies crosses the edges of the OPTOPUS fields.

\begin{figure*}
\epsfysize=16cm
%
\centerline{\rotate[l]{\epsfbox[28 110 423 785]{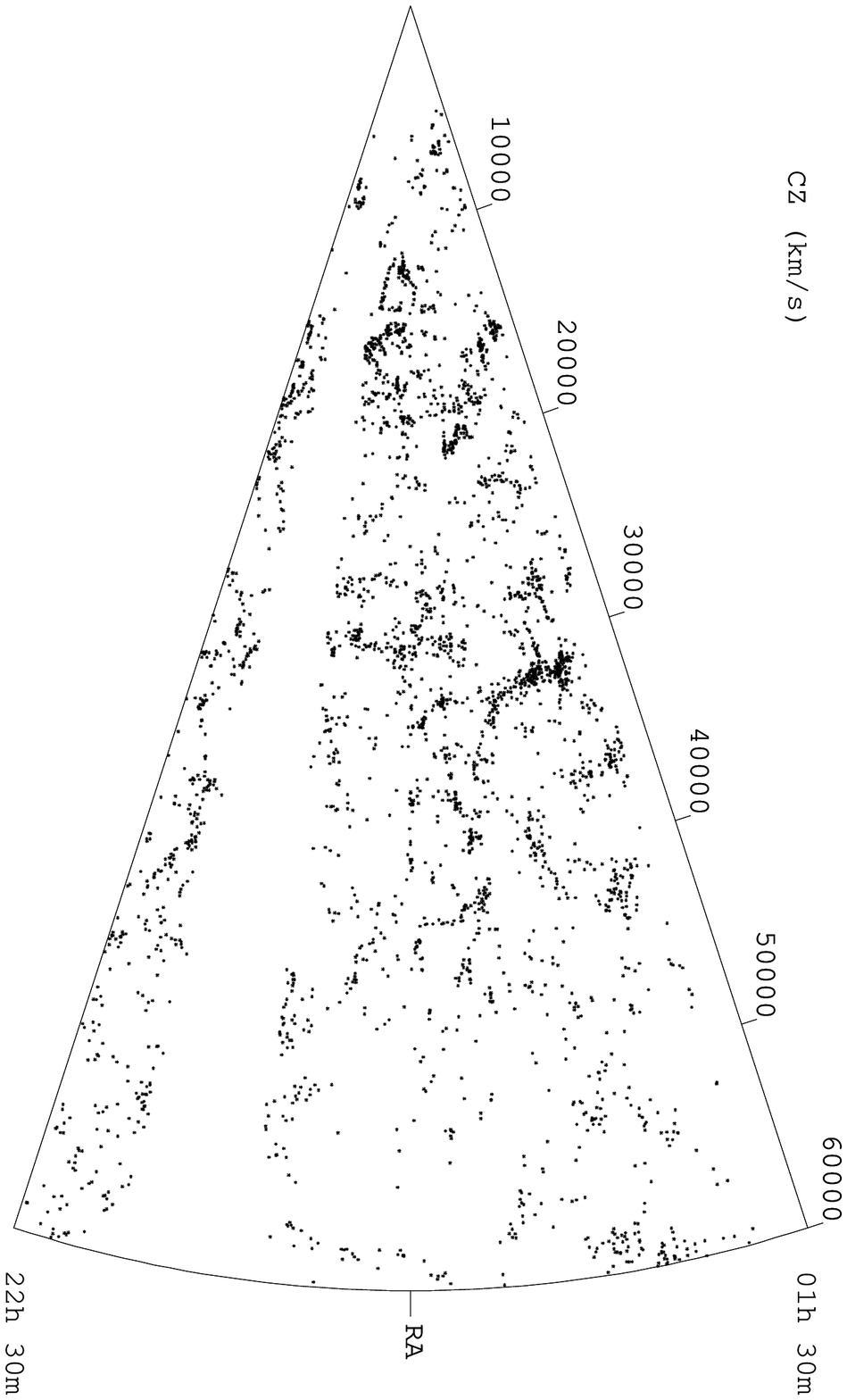}}}
\epsfysize=16cm
\centerline{\rotate[l]{\epsfbox[28 110 500 785]{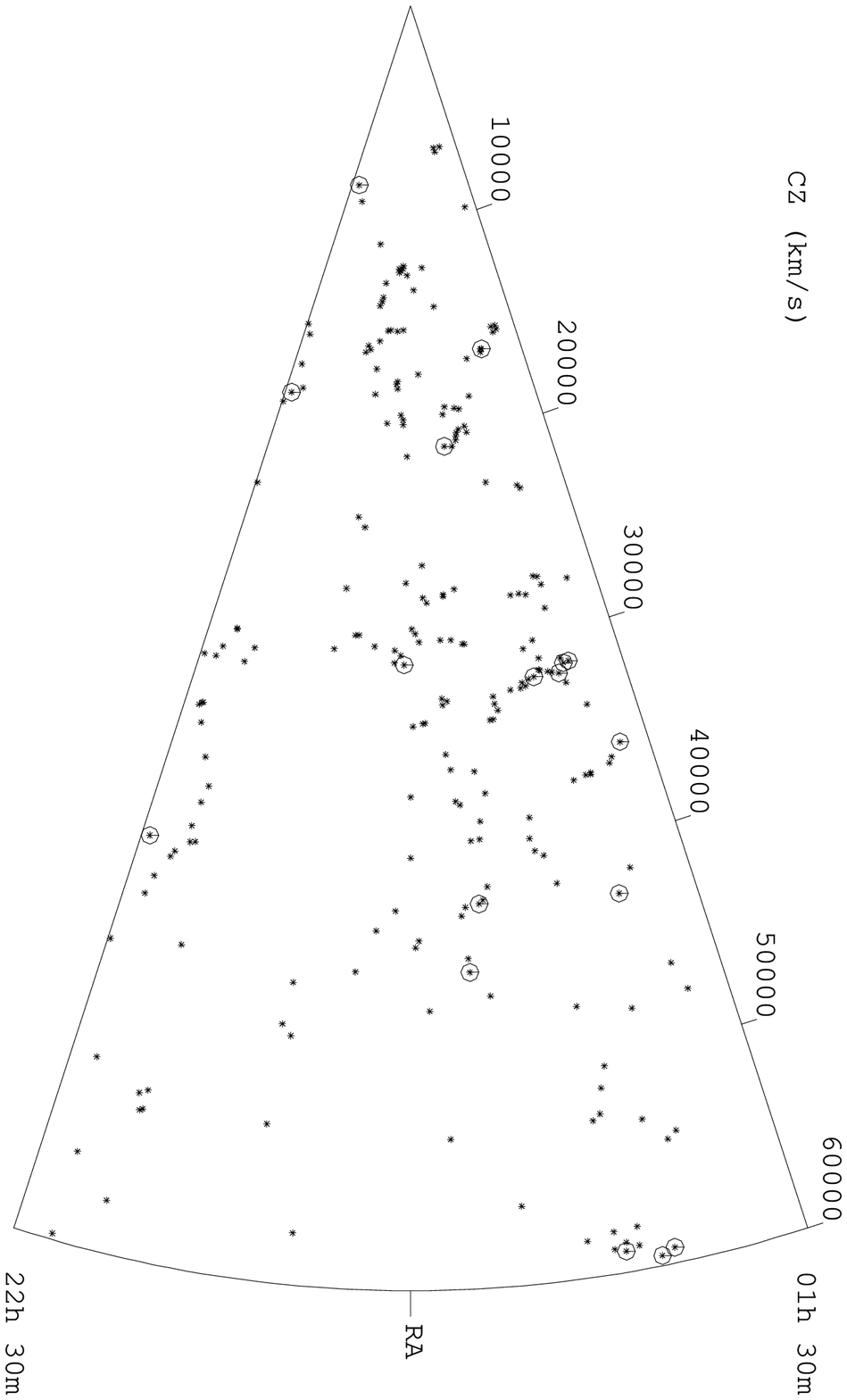}}}
\caption[]{Cone diagrams ($\alpha$ -- $cz$) of ESP galaxies (top panel)
and of ESP groups (bottom panel). The larger circles in the cone
diagram of groups mark the ESP counterparts of known ACO and/or
EDCC clusters.
}
\end{figure*}

\section{Properties of Groups} 
\medskip 

In this section we discuss properties of  ESP groups 
that  can be used to characterize the LSS and that set useful
constraints to the predictions of cosmological N-body models.

\medskip \medskip 

\subsection{Abundances of Groups and Members} 
\medskip

The first ``global''
property of groups we consider is  the ratio, $f_{groups}$, of their
number to the number of non-member galaxies within the survey.  For ESP
we have $f_{ESP,groups}$ = 0.13 $\pm$ 0.01, for CfA2N RPG find
$f_{CfA2N,groups}$ = 0.13 $\pm$ 0.01, for SSRS2 Ramella \etal 1998 find
$f_{SSRS2,groups}$ = 0.12 $\pm$ 0.01.  Clearly the proportion of groups
among galaxies is the same in all three independent volumes of the
universe surveyed with ESP, CfA2N and SSRS2. Because CfA2N and SSRS2
mostly sample only one large structure while ESP intercepts
several large structures, 
our result means that the clustering of
galaxies in groups within the large scale structure is homogeneous
on scales smaller than those of the structures themselves. 

We point out that, on the
basis of our simple simulation, we do not expect the OPTOPUS mask to
affect the determination of $f_{groups}$.

We now consider the ratio of member to non-member galaxies, $f_{mem}$.
Within ESP we have $f_{ESP,mem}$ = 0.68 $\pm$ 0.02;  within CfAN and
SSRS2, the values of the ratio are $f_{CfA2,mem}$ = 0.81 $\pm$ 0.02
and $f_{SSRS2,mem}$ = 0.67 $\pm$ 0.02 respectively.  Quoted uncertainties
are one poissonian standard deviation.  According to the poissonian
uncertainties, $f_{ESP,mem}$ and $f_{SSRS2,mem}$ are
undistinguishable.  The value of $f_{CfA2,mem}$ is significantly
different from the other two ratios.  However, the real
uncertainty in the ratio of members to non-members is higher than the
poissonian estimate because the fluctuations in the number of members 
is dominated by the fluctuations in the
smaller number of groups.  Moreover, the total number of members is
strongly influenced by few very rich systems. In fact, it is sufficient
to eliminate two clusters, Virgo and Coma, from CfA2N in order to
reduce the value of $f_{CfA2,mem}$ to $f_{CfA2,mem}$ = 0.70 $\pm$
0.02, in close agreement with the ratio observed within ESP and SSRS2.

In conclusion, groups are a remarkably stable property of the
large-scale distribution of galaxies. Once the richest clusters 
are excluded, the abundances of groups and of
members  relative to that
of non-member or``field'' galaxies are constant over several large and
independent regions of the universe.

\medskip \medskip
\subsection{ Distribution of Groups in Redshift-Space}
\medskip

We plot in the top panel of Figure 4 the cone diagram 
($\alpha$ vs $cz$) for the 3085 ESP
galaxies within 5000 $< cz <$ 60000 \kms.  In the bottom panel of
Figure 4 we plot the
cone diagram of the 231 ESP groups. 
Figures 4 shows that groups  trace very well the galaxy distribution, as
they do in shallower surveys ($cz \simless$ 12000 \kms). Note that
in Figure 4 we project adjacent beams, not a strip
of constant thickness.

The topology of the galaxy distribution in redshift space has already
been described by Vettolani \etal (1997) and will be the subject of a
forthcoming paper.  The most striking features are the voids of sizes
$\simeq$ 50 \mpc \ and the two density peaks at $cz \simeq$ 18000 \kms
and  $cz \simeq $ 30000 \kms.  These features are also the main
features of the group distribution.

\begin{figure}
\epsfysize=9cm
\epsfbox{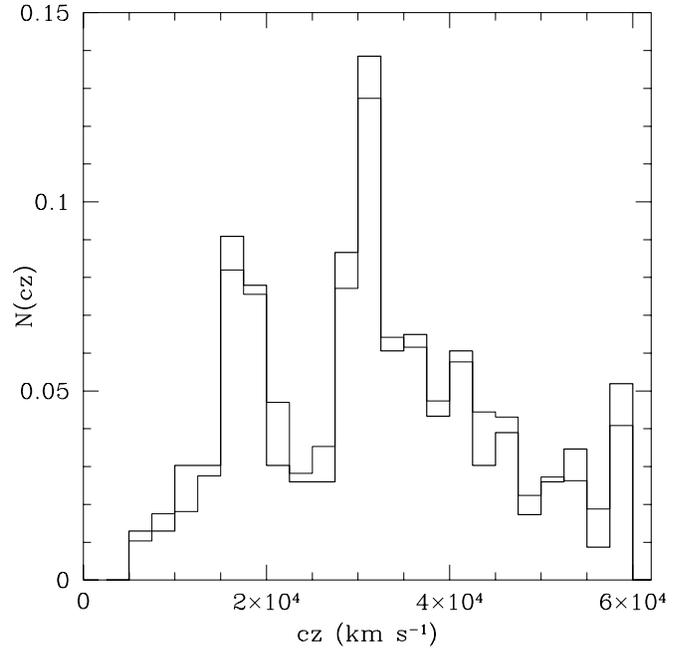}
\caption[]{The redshift distributions of groups (thick
histogram) and galaxies (thin histogram), divided by
the total number of groups and 
by the total number of galaxies  respectively.
}
\end{figure}

\begin{figure}
\epsfysize=13cm
\epsfxsize=9cm
\epsfbox{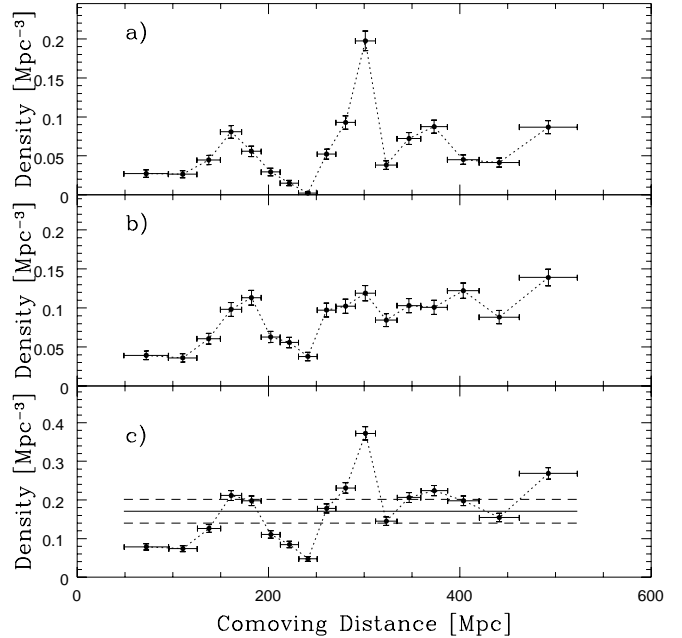}
\caption[]{Number densities of galaxies in comoving distance bins. 
The top panel
is for members, the middle panel is for non-members, and the bottom
panel is for all galaxies.The dashed lines represent the $\pm$ one sigma corridor around the mean
galaxy density.
}
\end{figure}

In Figure 5 we plot the  redshift distributions of groups (thick
histogram) and galaxies (thin galaxies), divided by the total number of
groups and by the total number of galaxies, respectively.  In Figure 6
we plot number densities of galaxies in redshift bins.  Number
densities are computed using the $n_3$ estimator of Davis \& Huchra (1982): 
all the details about the density estimates are given in Zucca \etal 1997. 
We vary the size of the redshift bins in order
to keep constant the number of galaxies expected in 
each bin based on the selection function.  The top panel is for 
member galaxies, the middle panel is
for isolated and binary galaxies, and the bottom panel is for all ESP galaxies.
The dashed lines represent the $\pm$ one sigma corridor around the mean
galaxy density.

It is clear from Figure 5 that the redshift distributions of groups and
galaxies are undistinguishable (98 \% confidence level).  Not
surprisingly, the number density in redshift bins of members and all
galaxies are highly correlated (Figure 6a and 6c).  More interestingly,
the number density distribution of non-member galaxies is also correlated
with  the number density distribution of all galaxies (Figure 6b and
6c).  In particular, the two density peaks at  $cz \simeq $ 18000
\kms and $cz \simeq $ 30000 \kms of the  number density distribution
of all galaxies are also identifiable in the number density
distribution of non-member galaxies, even if with a lower contrast.

We know from the previous section that groups are a very stable global
property of the galaxy distribution within the volume of the ESP and
within other shallower surveys. Here we show that a tight relation
between non-member galaxies and groups exists even on smaller scales. 

Our result is particularly interesting in
view of the depth of the ESP survey and of the number of large
structures intercepted along the line-of-sight.

\medskip \medskip 
\subsection{The Distribution of Velocity
Dispersions} 
\medskip

We now discuss the velocity dispersions of ESP groups. According to our
simulation in section 2, the effect on the
velocity dispersions of the OPTOPUS mask is statistically negligible.

The median velocity dispersion of all groups is $\sigma_{ESP,median}$ =
194 (106,339) \kms.  The numbers in parenthesis are the 1st and 3rd
quartiles of the  distribution.  
Poor groups with \nmem $<$ 5 have a median velocity dispersion
$\sigma_{median,poor}$=145 (65,254) \kms, richer groups have
$\sigma_{median,rich}$=272 (178,399) \kms.  For comparison, the median
velocity dispersions of CfA2N and SSRS2 are $\sigma_{CfA2N,median}$
=198 (88,368) \kms and $\sigma_{SSRS2,median}$ = 171 (90,289) \kms. We take the
values of the velocity dispersions for the CfA2 and SSRS2 groups from Ramella 
\etal (1997,1998). In order to compare these velocity dispersions with
ours, we correct them for a fixed
error of 50 \kms (corresponding to an RVSAO error of $\simeq$ 35 \kms)
and multiply them by $\sqrt{(N_{mem}-1)/(N_{mem}-3/2)}$. We note that,
because of the OPTOPUS mask, the comparison of the velocity dispersions of
"rich" and "poor" groups within ESP with those of similar systems
within CfA2N and SSRS2 is not meaningful. A fraction of ESP 
"poor" groups may actually be part of "rich" groups.

In Figure 7 we plot (thick histogram) the distribution of the velocity
dispersions, $n_{ESP}(\sigma)$, normalized to the total number of
groups. Errorbars are one sigma poissonian errors.  We also plot the
normalized \sv distributions of CfA2N and SSRS2.  
According to the KS test, differences  between
$n_{ESP}(\sigma)$ and the other two distributions are not significant 
(P$_{KS}$ = 0.3 and 0.2 for the comparison between ESP and CfA2N and
SSRS2 respectively).

\begin{figure}
\epsfysize=9cm
\epsfbox{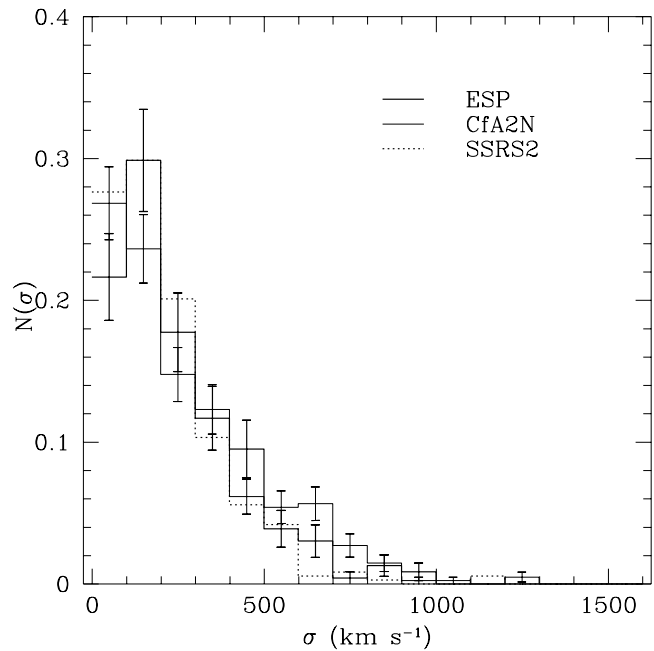}
\caption[]{Comparison between the distribution of velocity dispersions
of ESP groups (thick line) and those of CfA2N (thin line) and SSRS2 groups
(dotted line).
Each distribution is normalized to the total number of
groups. Errorbars are one sigma poissonian errors.
}
\end{figure}

It is interesting to point out that $n_{CfA2N}(\sigma)$ and
$n_{SSRS2}(\sigma)$ do differ significantly (97\% level),
$n_{CfA2N}(\sigma)$ being richer of high velocity dispersion systems
(Marzke \etal, 1995). Groups with  dispersion velocities \sv $>$ 700
\kms are rare and the fluctuations from
survey to survey correspondingly high. The abundance of these high \sv
systems is the same within both ESP and SSRS2 (2\%) but it is higher
within CfA2N (5\%).  If we disregard these few high velocity dispersion
systems, the difference between $n_{CfA2N}(\sigma)$ and
$n_{SSRS2}(\sigma)$ ceases to be significant.
From this result we conclude that each survey contains
a fair representation of groups.

The distribution of velocity dispersions is an important characteristic
of groups because it is linked to the group mass. Therefore $n(\sigma)$
constitutes an important constraint for cosmological models.  
Furthermore, \sv is a
much better parameter for the classification of systems than the number
of members (even more so in the case of the present catalog where the
OPTOPUS mask affects the number of members much more than velocity
dispersions)

The ESP survey provides a new determination of the shape of $n(\sigma)$
in a much deeper volume than those of existing shallower surveys. We
find that, within the errors, $n_{ESP}(\sigma)$  is very similar to
both $n_{CfA2N}(\sigma)$ and $n_{SSRS2}(\sigma)$.

\medskip \medskip 

\section{Properties of Galaxies in Groups}
\medskip
In this section we examine the luminosities and the spectral features
of galaxies in different environments: the ``field'', groups, and clusters.
The dependence of these properties on the environment offers insights 
into the processes of galaxy formation and evolution and on the
dynamical status of groups. 

\subsection{The Luminosity Function of Members}
\medskip

Here we investigate the possible difference between the luminosity functions
of member and non-member galaxies.
We compute the luminosity function with the STY
method (Sandage, Tamman \& Yahil 1979). We assume a Schechter (1976) form
for the luminosity function and follow the procedure described 
in detail in Zucca \etal (1997).

We find that galaxies in groups have a brighter $M^*$ with respect to 
non--member galaxies; the slope $\alpha$ does not change significantly
in the two cases. In particular the parameters we obtain 
are $\alpha= -1.25^{+0.11}_{-0.11}$ and $M^* = -19.80^{+0.14}_{-0.13}$ 
for the 1250 members, and  $\alpha= -1.21^{+0.10}_{-0.09}$ 
and $M^* = -19.52^{+0.10}_{-0.10}$ for the 1835 non--members. 

In Figure 8 we draw (dotted lines) the confidence ellipses of the $\alpha$
and $M^*$ parameters obtained in the two cases of member and non-member
galaxies.  The two luminosity functions differ at the
$2\sigma$ level.  In Figure 8 we also plot the confidence
ellipses for the parameters of the total sample (solid lines) derived
in the same volume of ESP where we identify groups.

\begin{figure}
\epsfysize=9cm
\epsfbox{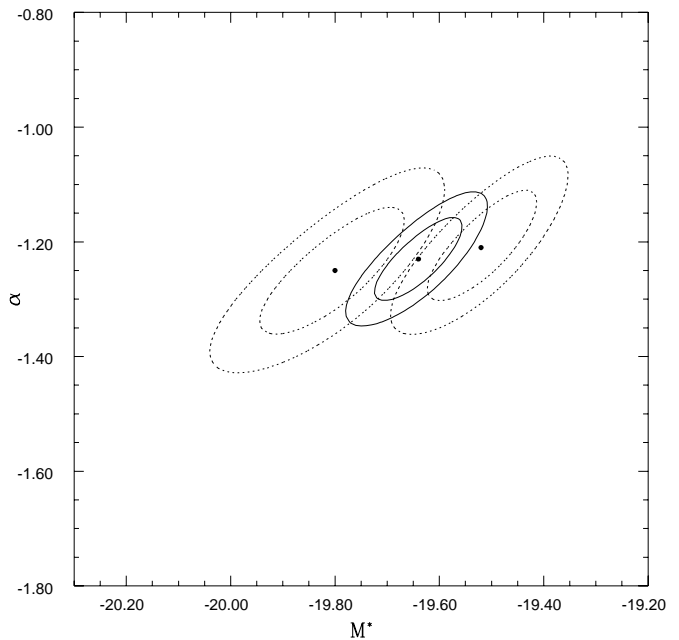}
\caption[]{One- and two-sigma 
confidence ellipses  (dotted lines) of the $\alpha$
and $M^*$ parameters of the Schechter luminosity function
of members (brightest $M^*$) and non-members (faintest $M^*$).
The solid lines show the confidence ellipses derived for the total ESP sample 
considered in this paper
(5000 \kms $\le$ \cz $\le$ 60000 \kms).
}
\end{figure}

The fact that galaxies in groups are brighter than non-member galaxies is
a clear demonstration of the existence of luminosity segregation
in the ESP survey, a much deeper survey than those where the
luminosity segregation has been previously investigated 
(Park \etal 1994, Willmer \etal 1998). Our finding is consistent
with the results of Lin \etal (1996), who find evidence 
of a luminosity bias in their analysis of the LCRS power spectrum.

In further support of the existence of a luminosity segregation, 
we also find that  $M^*$ becomes brighter for members
of groups of increasing richness. As before, the parameter 
$\alpha$ remains almost constant. Only in the case of the richest groups,
$N_{mem}\ge 10$, we find a marginally significant steepening 
of the slope $\alpha$.

\medskip 
\subsection{Emission/Absorption Lines
Statistics} \medskip

One interesting question is whether
the environment of a galaxy has a statistical
influence on the presence of detectable emission lines in the galaxy spectrum.
Because emission lines galaxies are mostly spirals (Kennicutt 1992), 
the answer to this question is relevant to the investigation of 
the morphology-density relation in systems of intermediate density.

The fraction of ESP galaxies with detectable emission lines within the
redshift range $5000 \le cz \le 60000$ \kms is 44\% (1360/3085). Of
these \egals, (34 $\pm$ 2)\% (467/1360) are members of groups. The fraction of
galaxies without detectable emission lines, \agals, that are members of
groups is significantly higher: 783/1725 or (45 $\pm$ 2)\%. 
We note that our detection limit for emission lines correspond to 
an equivalent width of about 5 $\AA$

\begin{figure}
\epsfysize=9cm
\epsfbox{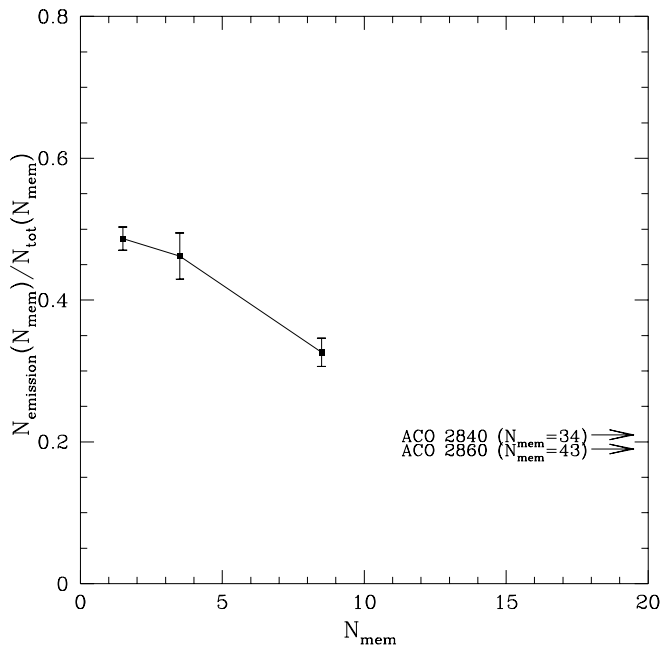}
\caption[]{Fraction of \egals in the ``field'', in poor groups, and rich groups.  The
two arrows indicate the fraction of \egals in the two richest ACO
clusters in our catalog, A2840 ($f_e$ = 21\%) and A2860 ($f_e$ = 19\%)
}
\end{figure}

We consider three types of environments: a) the ``field'',
i.e.  all galaxies that have not been assigned to groups, b) poor
goups, i.e. groups with  3 $\le$ \nmem $\le$ 4, and c) rich groups 
with 5 $\le$ \nmem. We find that the fraction of \egals decreases as
the environment becomes denser. In the ``field'' the fraction of \egals is
$f_e$ = 49\%, it decreases to $f_e$ = 46\%  for poor groups and to 
$f_e$ = 33\% for richer groups.  In Figure 9 we plot $f_e$ as a function of \nmem. We also
indicate the values of $f_e$ of the two richest Abell clusters in our
catalog, A2840 ($f_e$ = 21\%) and A2860 ($f_e$ = 19\%). 

The significance of the correlation between environment and $f_e$
can be investigated with a 2-way contingency table (Table 2). For simplicity,
we do not consider triplets and quadruplets.

\setcounter{table}{1}
\begin{table}
\caption[]{ Frequency of Emission Line Galaxies in Different Environments}
\begin{flushleft}
\begin{tabular}{lrrr}
\hline\noalign{\smallskip}
  & $N_{e.l.}$ & $N_{a.l.}$ &  $N_{tot}$ \cr 
\noalign{\smallskip}
\hline\noalign{\smallskip}
  ``field''         &  893  & 942  & 1835 \\
  rich groups   &  266  & 549  &  815 \\
  total         & 1159  &1491  & 2650 \\
\noalign{\smallskip}
\hline
\end{tabular}
\end{flushleft}
\end{table}

The contingency coefficient is C=0.15 and the significance of
the correlation between environment and frequency of emission line
galaxies exceeds the 99.9\% level.


Our result indicates that the  morphology-density relation extends over the
whole range of densities from 
groups to clusters. Previous indications of the existence of the
morphology-density relation for groups are based either on very
local samples (Postman \& Geller 1984) or on samples
that are not suitable for statistical analysis (Allington-Smith \etal 1993).
Very recently, Hashimoto \etal (1998) also confirm the existence of a
morphology-density relation over a wide range of environment densities
within LCRS.

Examining our result in more detail,
we note that the fraction of  \egals, $f_e$, in triplets and
quadruplets is very similar to the value of $f_e$ for isolated galaxies. 
Triplets and quadruplets are likely to correspond, on average,
to the low-density tail of groups. Moreover,
Ramella \etal (1997) and Frederic (1995a) estimate that the FOFA 
could produces a large fraction of unbound triplets and quadruplets.
These ``pseudo-groups''  dilute the properties of real bound triplets
and quadruplets with ``field'' galaxies, 
artificially increasing the value of $f_e$. This effect, in our survey,
is partially counter-balanced by the triplets and quadruplets that are
actually part of richer systems cut by the OPTOPUS mask. Considering that
rich systems are significantly rarer than triplets and quadruplets, we
estimate that the value of $f_e$ we measure for triplets and quadruplets
should be considered an upper limit.

Our catalog also includes ESP counterparts of 17 clusters listed in at
least one of the ACO, ACOS (Abell \etal 1989) 
or EDCC (Lumsden \etal 1992) catalogs (section 8, Table 3).
For these clusters $f_{e,clusters}$ = 0.25 (63 \egals out of 256
galaxies). The number of members of these systems
is not a direct measure of their richness because of the apparent
magnitude limit of the catalog and because of the OPTOPUS mask.
However, because they include all the richest systems in our catalog
and  because they are counterparts of 2-D clusters, it is  resonable to
assume that they are intrinsically rich. We remember here that Biviano
\etal  (1997) find $f_e$ = 0.21 for their sample of ENACS clusters.The fact
that for ESP counterparts of clusters we find a lower value of $f_e$
than for the other rich groups ($f_{e,groups}$ = 0.36 without clusters),
further supports
the existence of a morphology-density relation over the whole range
of densities from clusters to the ``field''.

Many systems of 
our catalog are not completely surveyed, therefore
the relationship between $f_e$ and the density of the environment we
find can only be considered qualitative. However, while incompleteness
certainly increases the variance around the mean result, we do not
expect severe systematic biases. In
order to verify that incompleteness does not  affect our
results, we consider the subsample of 67 groups that contain no ESP
objects without measured redshift. We  obtain for this subsample the
same relationship between group richness and fraction of emission line
galaxies we find for the whole catalog.

\medskip 
\subsection{Seyfert Galaxies} \medskip

Within ESP we identify 12 Seyfert 1 galaxies and 9 Seyfert 2 galaxies.
We identify type 1 Seyferts  visually on the basis of the
presence of broad (FWHM of a few $10^3$~km~s$^{-1}$) components in the
permitted lines. Our list is complete with the possible exception of
objects with weak broad lines which are hidden in the noise.

The identification of type 2 Seyferts is not straightforward, because
it is based on line ratios and usually requires measurements of
emission lines which fall outside our spectral range: only the
F([O~III]$\lambda$5007)/F(H$\beta$) ratio is available from our
spectra, and it is therefore impossible to draw a complete diagnostic
diagram (Baldwin \etal 1981, Veilleux \& Osterbrock 1987).  We 
classify tentatively as  type 2 Seyferts all emission line galaxies with
$\log(\hbox{F([O~III]$\lambda$5007)}/\hbox{F(H$\beta$)})\ge 0.7$: this
threshold cuts out almost all non-active emission line galaxies, but
also many narrow-line AGN with a medium to low degree of ionization.
Thus the list of possible Seyfert 2 galaxies is almost free of
contamination, but should by no means be considered complete.

The origin of the Seyfert phenomenon could be related to the
interaction with close companions (Balick \& Heckman 1982, Petrosian
1982, Dahari 1984, MacKenty 1989), or to a dense environment
(Kollatschny \& Fricke 1989, De Robertis \etal 1998). Observational
evidence is, however, far from conclusive. For example, Seyfert 1 and
Seyfert 2 galaxies have been found to have an excess of (possibly)
physical companions compared to other spiral galaxies by Rafanelli
\etal (1995).  Laurikainen \& Salo (1995) agree with Rafanelli
\etal (1995) about Seyfert 2 galaxies, but reach the opposite
conclusion about Seyfert 1 galaxies.

In our case, 7 (33\%) out of 21 Seyferts are group members. For
comparison, 460 (34\%) emission line galaxies (not including Seyfert
galaxies) are group members and 879 are either isolated or binaries.
Clearly, within the limits of our relatively poor statistics, we find
that Seyfert galaxies do not prefer a different environment than that
of the other emission line galaxies.

In order to test the dependence of the Seyfert phenomenon on the
interaction of galaxies with close companions rather than with the
general environment, we compute for all Seyferts and emission line
galaxies the projected linear distance to their nearest neighbor, the
$nn$-distance. We limit the search of companions to galaxies that are
closer than 3000 \kms along the line of sight.

We find that the distribution of   $nn$-distances 
of the sample of Seyfert galaxies is not significantly different from
that of all emission line galaxies. 

We also consider the frequency of companions at projected linear
distances $d < 0.15 h^{-1}$ Mpc. We have 7 Seyfert galaxies with
such a close companion (33\%)  and 315 (23\%) emission lines galaxies.
One of the 7 Seyferts  is a member of a binary systems, the remaining
six Seyferts are members of groups.  Even if, taken at face value, the
higher frequency of close companions observed among Seyfert galaxies
supports  a causal connection between gravitational interaction and the
Seyfert phenomenon, these frequencies are not significantly different.


We note that members of close angular pairs ($\theta < 24.6$
arcsec) in the original target list for ESP, are more frequently missing
from the redshift survey than other objects
(Vettolani \etal 1998). This bias, due to OPTOPUS mechanical constraints,
could hide a possible excess of
physical companions of Seyfert galaxies.

In order to estimate how strongly our result could be affected by this
observational bias, we identify the nearest neighbors of 
Seyfert and emission line galaxies from a list including both 
galaxies with redshift and objects
that have not been observed. When we compute projected linear distances
to objects that have not been observed, we assume that they are at the
same redshift of their candidate companion galaxy. As before, we do not 
find significant differences between the new $nn$-distributions of
Seyferts and \egals.

This result demonstrates that the higher average incompleteness of
close angular pairs does not affect our main conclusions: a) Seyfert
galaxies within ESP are found as frequently within groups as other
emission line galaxies, b) Seyfert galaxies show a small but not significant
excess of close physical companions relative to the other emission line
galaxies. We point out again that the sample
of Seyferts is rather small and the statistical uncertainties
correspondingly large.

\medskip \medskip 
\begin{table*}
\caption[]{ Clusters within ESP}
\begin{flushleft}
\begin{tabular}{crrcccccc}
\hline\noalign{\smallskip}
 ID & ESP & \nmem & $\alpha_{1950}$ & $\delta_{1950}$ & R & $z_{est}$ & $z_{ESP}$& $\sigma$  \\ 
    &    &     & ($^h~^m~^s$)   &($^o$~'~'')    &   &    &    & \kms  \\ 
\noalign{\smallskip}
\hline\noalign{\smallskip}
 E0163 &   6 &   3 & 22~32~40 & -40~38~41 &  &         &0.13535& 121 \\
 E0169 &   7 &  11 & 22~34~12 & -39~50~51 &  &         &0.06294& 282 \\
 S1055 &  21 &   9 & 22~39~43 & -40~16~07 &0 &         &0.02901& 102 \\
 A4068 &  88 &   9 & 23~57~08 & -39~46~59 &0 & 0.07151 &0.10261& 700 \\
 E0435 & 121 &   8 & 00~17~31 & -40~40~37 &  &         &0.15073& 334 \\
 A2769 & 126 &   6 & 00~21~45 & -39~53~49 &0 & 0.15708 &0.14020& 419 \\
 A2771 & 128 &  18 & 00~21~50 & -40~26~49 &0 & 0.06260 &0.06876& 268 \\
 A2828 & 176 &   5 & 00~49~10 & -39~50~54 &0 & 0.13133 &0.19676& 468 \\
 A2840 & 183 &  34 & 00~52~01 & -40~04~19 &1 & 0.10460 &0.10618& 339 \\
 A2852 & 192 &  12 & 00~57~00 & -39~54~19 &0 & 0.17581 &0.19845& 235 \\
 E0113 & 196 &  16 & 00~58~21 & -40~31~05 &  &         &0.05449& 372 \\
 A2857 & 200 &   8 & 01~00~06 & -40~12~42 &1 & 0.19092 &0.19755& 504 \\
 E0519 & 205 &  43 & 01~02~36 & -40~03~02 &  &         &0.10637& 319 \\
 S0127 & 213 &  25 & 01~05~27 & -40~21~08 &0 &         &0.10498& 505 \\
 A2874 & 216 &  25 & 01~06~08 & -40~36~01 &1 & 0.15812 &0.14191& 817 \\
 E0529 & 218 &  17 & 01~07~40 & -40~40~34 &  &         &0.10483& 282 \\
 E0546 & 231 &   7 & 01~19~27 & -39~53~07 &  &         &0.11909& 424 \\
\noalign{\smallskip}
\hline
\end{tabular}
\end{flushleft}
\end{table*}

\section{Clusters and Rich Systems} \medskip

Within our survey lie the centers of 9 ACO clusters,
5 ACOS clusters  and 12 EDCC  clusters.
Several entries of the three lists correspond to the same cluster.
Taking into account multiple identifications, there are 20 
clusters listed within one or more of the three catalogs that
lie within ESP.

In our catalog we find at least one counterpart for 17 out of the 20
clusters. The three clusters that do not correspond to any of our
systems are ACO 2860, ACOS 11, and ACOS 32, all of Abell richness R =
0.  ACO2860 is a very nearby object with a redshift, $z$ = 0.0268,
close to our minimum redshift. ACOS 11 and ACOS 32 are both distance
class D = 6 objects that may be either projection effects or real
clusters located beyond our redshift limit.  We select the ESP
counterparts among the groups that are close to the clusters on the sky
and that have a redshift compatible with the distance class and/or
magnitude of the cluster.  If more ESP groups are counterparts of a
cluster, we identify the cluster with the richest counterpart.

In Table 3 we list the name (column 1), and the coordinates (columns 4,
5) of the 17 clusters with ESP counterpart together with their richness
(column 6) and, if available, their redshift as estimated by Zucca
\etal (1993) (column 7).  In the case of clusters listed in both EDCC and ACO
or ACOS, we give the ACO or ACOS identification number.  In the same
Table 3 we also list the ID number of the cluster counterparts within
our catalog (column 2), their number of members (column 3), redshift
(column 8) and velocity dispersion (column 9).

There are 8 clusters with redshift estimated by Zucca \etal (1993) The
measured redshifts of 6 of these clusters are in good agreement with
the estimated redshift: the difference between the two redshifts is of
the order of 10\%, less than the 20\% uncertainty on the estimated
redshifts. For the remaining 2 clusters, ACO 2828 and ACO 4068, the
estimated redshift is significantly smaller than our measured redshift.
The projection of the foreground systems  ESP 175 and ESP 178 within
the Abell radius of ACO 2828 could explain the inconsistency between
estimated and measured redshift for this cluster. In the case of ACO
4068 we do not find any foreground/background system within ESP. ACO
4068 is very close to the northern declination boundary of the ESP
strip. An inspection of the COSMOS catalog just outside the boundary of
the OPTOPUS field containing ACO 4068 shows that a significant part of
this cluster lies outside our redshift survey and therefore
background/foreground projection could still be the cause of the
inconsistency between its estimated and measured redshifts.

We also note that EDCC163 and ACOS1055 are among the most incomplete
systems in our catalog. In the fields of EDCC163 (\nmem = 3) and ACOS1055 
(\nmem = 9) the number of  objects without redshift is 
16 and 63  respectively.
We will not consider these two clusters in what follows.

In panel a) and panel b) of Figure 10 we plot , respectively, \nmem and
\sv as a  function of \cz. As expected, clusters (represented by large
dots) populate the highest part of both diagrams at all redshifts. In
both diagrams, mixed with clusters, there are also ESP groups
that have not been identified as clusters.

\begin{figure}
\epsfysize=9cm
\epsfbox{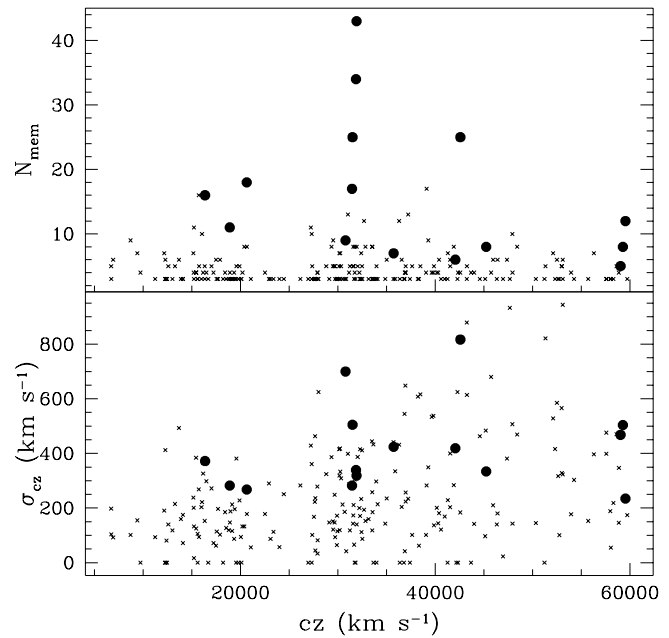}
\caption[]{\nmem (top panel) and \sv (bottom panel) as a function of \cz.
Large dots are ESP counterparts of 2-D ACO and/or EDCC catalogs.
}
\end{figure}

\begin{figure}
\epsfysize=9cm
\epsfbox{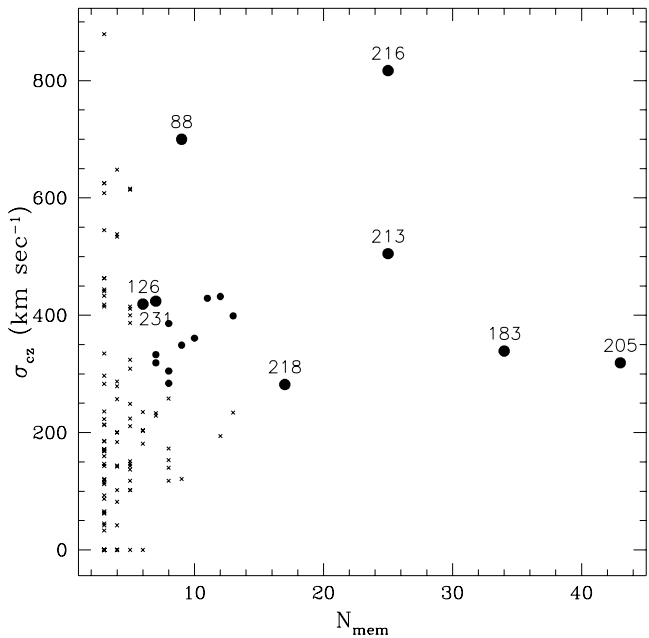}
\caption[]{ ESP groups (crosses) with 25000 \kms 
$<$ \cz $<$ 45000 \kms
in the \sv -- \nmem plane.
Large dots are ESP counterparts of 2-D ACO and/or EDCC catalogs,
smaller dots are ``cluster-like'' groups. The ESP counterparts 
of ACO and/or EDCC clusters are labeled with their ESP ID number (Table 1).
}
\end{figure}

The completeness of bidimensional cluster catalogs is an important
issue for cosmology (van Haarlem \etal 1997) since the density of these
clusters and their properties are used as constraints on cosmological
models (e.g. Frenk \etal 1990, Bahcall \etal 1997, Coles \etal 1998).
It is therefore interesting to determine whether there are other
systems selected in redshift space that have properties similar to
those of the  cluster counterparts but that have escaped 2-D
identification.

We limit our search for "cluster-like" groups to the velocity range
25000 $<$ \cz $<$ 45000 \kms. Within this range the selection function
is rather stable and relatively close to its maximum. In our catalog 
we identify the counterparts of 8 2-D clusters within this redshift range. 
Two of the eight clusters are of richness class R=1 
(ACO2840=ESP183 and ACO2874=ESP216). 
The minimum number of members of these clusters is 6 
and the lowest  velocity dispersion of the 8 clusters 
is  280 \kms. 

Apart from the counterparts of the 8 clusters, we find 11 additional ESP groups
that satisfy all three conditions 25000 $<$ \cz $<$ 45000 \kms,
$N_{mem,min} \geq 6$, and $\sigma_{cz,min} \geq 280$ \kms. These groups are $\simeq$ 10\% of all groups in this redshift interval and
we list them in Table 4.

\begin{table}
\caption[]{ Cluster-like Groups }
\begin{flushleft}
\begin{tabular}{rrcccc}
\hline\noalign{\smallskip}
  ESP &  \nmem& $ \alpha_{1950}$ & $ \delta_{1950}$ & $cz$ & $ \sigma$  \\ 
  &     & $h~~m~~s$ & $^o~~`~~"$  & \kms & \kms \\  
\noalign{\smallskip}
\hline\noalign{\smallskip}
    48 &  7& 23~26~03 & -39~52~45 & 30231&   333 \\
    96 &  8& 00~02~04 & -40~29~17 & 29329&   386 \\
   124 & 17& 00~20~33 & -40~33~37 & 39102&   283 \\
   130 &  8& 00~23~02 & -40~15~47 & 41898&   284 \\
   155 &  9& 00~40~39 & -40~00~46 & 39280&   349 \\
   186 &  7& 00~54~16 & -40~25~00 & 30163&   319 \\
   190 & 13& 00~55~29 & -40~34~41 & 31033&   399 \\
   195 &  8& 00~58~10 & -40~06~00 & 31728&   305 \\
   201 & 11& 01~00~34 & -40~03~30 & 27225&   429 \\
   203 & 10& 01~02~28 & -40~18~29 & 27298&   361 \\
   226 & 12& 01~14~57 & -39~52~45 & 36298&   432 \\
\noalign{\smallskip}
\hline
\end{tabular}
\end{flushleft}
\end{table}

In a \sv -- \nmem plane, Figure 11, the eleven "cluster-like" groups 
occupy a "transition region" between clusters and groups.
First we note that, in this plane, the two counterparts of the R=1 ACO clusters 
(ESP183 and ESP216) are very distant from the "cluster-like" groups.
The same holds true for the only rich EDCC cluster that is not an ACO cluster,
EDCC519. We conclude that no rich cluster is missing from 2-D catalogs in
the region of the sky covered by the ESP survey.
This conclusion is reassuring, even if it does not allow us to
discuss the problem of the completeness of rich 2-D clusters in general
because it is based on a small number of objects.

In the case of the more numerous poorer clusters, Figure 11 shows that
several systems could be missing from the 2-D list. 
The  boundaries of the cluster and group regions in the \sv -- \nmem plane
are blurred by the OPTOPUS mask and by the 
narrow width of the ESP survey. It is
therefore difficult to give a precise estimate of
how many "cluster-like" groups should be considered "missing" 
from bidimensional catalogs.

That poor 2-D clusters and  "cluster-like" 3-D groups are probably
the same kind of systems is confirmed by the fact that they have
the same fraction of \egals, a higher value
than it is typical of richer clusters.
The 11 "cluster-like" groups have a total of 110 members,
43 of which are \egals: $f_{e,cluster-like}$ = 0.39. The 4 poor
clusters that have \nmem $\le $ 17 include
39 members and have $f_{e,poor~clusters}$ = 0.41.
We remember here that for all ESP counterparts of clusters we find
$f_{e,clusters}$ = 0.25. 

In conclusion, the comparison of ESP systems with  ACO, ACOS and EDCC
clusters indicates that the ``low mass'' end of the distribution 
of clusters is poorly represented in 2-D catalogs; on the other hand,
the 2-D catalogs appear reasonably complete for high mass clusters. 

\section{Summary}

In this paper we search objectively and analyze groups of galaxies in
the recently completed ESP survey ($23^{h} 23^m \le \alpha_{1950} \le 01^{h} 20^m $ and  $22^{h} 30^m \le \alpha_{1950} \le
22^{h} 52^m $; $ -40^o 45' \le \delta_{1950} \le -39^o 45'$).
We identify 231 groups above the number overdensity threshold \drho=80
in the redshift range 5000 \kms $\le cz \le $ 60000 \kms. 
These groups contain 1250 members, 40.5 \% of the 3085 ESP
galaxies within the same \cz range. The median velocity dispersion of  
ESP groups is $\sigma_{ESP,median}$ = 194 \kms (at the redshift
of the system and taking into account measurement errors). We verify that
our estimates of the average velocity dispersions are not biased by the
geometry of the ESP survey which causes most systems to
be only partially surveyed. 

The groups we find trace very well the geometry of the large scale
distribution of galaxies, as they do in
shallower surveys. Because groups are also numerous, 
they constitute an interesting characterization of the large scale structure.
The following are our main results on the properties of groups
that set interesting ``constraints'' on cosmological models:

\begin{itemize}

\item   the ratio of members to non-members is $f_{ESP,mem}$ 
= 0.68 $\pm$ 0.02. This value is in very close agreement with the value 
found in shallower surveys, once the few richest clusters 
(e.g. Coma and Virgo) are neglected.

\item the ratio of groups to the number of non-member galaxies is
$f_{ESP,groups}$ = 0.13 $\pm$ 0.01, also in very close agreement with the value 
found in shallower surveys.

\item the distribution of velocity dispersions of ESP groups
is not distinguishable from those of CfA2N and SSRS2  
groups.

\end{itemize}

These results are of particular interest because 
the ESP group catalog is five times deeper than any other wide-angle
shallow survey group catalog and the number
of large scale features explored is correspondingly larger. As a consequence,
the properties of ESP groups are more stable with respect to 
possible structure-to-structure variations. The fact that 
the properties of ESP groups agree very well with those
of CfA2N and SSRS2 groups indicates that structure-to-structure
variations are not large and that the properties of groups we find
can be considered representative of the local universe.

As far as the richest systems (clusters) are concerned, 
we identify ESP counterparts for 17 out of 20 2-D 
selected ACO and/or EDCC clusters. Because the volume
of ESP is comparable to the volume of individual shallower surveys, it is
not big enough to include a fair sample of
clusters. The variations from survey to survey in the number and
properties of clusters are large.
 
Turning our attention to properties of galaxies 
as a function of their environment, we find
that:

\begin{itemize} 

\item the Schechter luminosity function of galaxies in groups 
has a brighter $M^*$ (-19.80) with respect to 
non--member galaxies ($M^* = -19.52)$; the slope $\alpha$ ($\simeq 1.2$)
does not change significantly between the two cases.

\item $M^*$ becomes brighter for members
of groups of increasing richness. The parameter 
$\alpha$ remains almost constant; only in the case of the richest groups
we find a marginally significant steepening of the slope $\alpha$.

\item 34\% (467/1360) of ESP galaxies with
detectable emission lines are members of groups. The fraction of
galaxies without detectable emission lines in groups is significantly
higher: 45\% (783/1725). 

\item  the fraction of \egals in the field is
$f_e$ = 49\%; it decreases to $f_e$ = 46\%  for poor groups and to 
$f_e$ = 33\% for richer groups. For the ESP counterparts of ACO and/EDCC
clusters $f_e$ = 25\%. 

\end{itemize} 

We conclude that luminosity segregation is at work
in the ESP survey: galaxies in the dense environment of 
groups are, on average, brighter than ``field'' galaxies.
Galaxies in groups are also less likely to have detectable emission lines
in their spectra. In fact, we find  a  gradual decrease of the
fraction of emission line galaxies among members of systems of
increasing richness: the  morphology-density relation clearly extends over the
whole range of densities from groups to clusters.

As a final note, we identify 12 Seyfert 1 galaxies and 9 Seyfert 2 galaxies.
We find that: a) Seyfert
galaxies within ESP are members of groups  as frequently  as other
emission line galaxies, and b) Seyfert galaxies show a small but not significant
excess of close physical companions relative to the other emission line
galaxies. We point out again that the sample
of Seyferts is rather small and the statistical uncertainties
correspondingly large.


\begin{acknowledgements}
We thank the referee for his careful reading of the manuscript and his
helpful suggestions.  This work has been partially supported through
NATO Grant CRG 920150, EEC Contract ERB--CHRX--CT92--0033, CNR Contract
95.01099.CT02 and by Institut National des Sciences de l'Univers and
Cosmology GDR.

\end{acknowledgements}



\end{document}